# Examining a solar–climate link in diurnal temperature ranges

Benjamin Laken[1,2,*], Jasa Čalogović[3], Tariq Shahbaz[1,2], and Enric Pallé[1,2]

[1] Instituto de Astrofísica de Canarias, Via Lactea s/n, E-38205, La Laguna, Tenerife, Spain
[2] Department of Astrophysics, Faculty of Physics, Universidad de La Laguna, Tenerife, Spain
[3] Hvar Observatory, Faculty of Geodesy, University of Zagreb, Kačićeva 26, HR-10000, Zagreb, Croatia
* Corresponding author: blaken@iac.es

**Abstract**
A recent study has suggested a link between the surface level diurnal temperature range (DTR) and variations in the cosmic ray (CR) flux. As the DTR is an effective proxy for cloud cover, this result supports the notion that widespread cloud changes may be induced by the CR flux. If confirmed, this would have significant implications for our understanding of natural climate forcings. Here, we perform a detailed investigation of the relationships between DTR and solar activity (total solar irradiance and the CR flux) from more than 60 years of NCEP/NCAR reanalysis data and observations from meteorological station data. We find no statistically significant evidence to suggest that the DTR is connected to either long-term solar periodicities (11 or 1.68-year) or short-term (daily-timescale) fluctuations in solar activity, and we attribute previous reports on the contrary to an incorrect estimation of the statistical significance of the data. If a CR–DTR relationship exists, based on the estimated noise in DTR composites during Forbush decrease (FD) events, the DTR response would need to be larger than 0.03°C per 1% increase in the CR flux to be reliably detected. Compared with a much smaller rough estimate of -0.005°C per 1% increase in the CR flux expected if previous claims that FD events cause reductions in the cloud cover are valid, we conclude it is not possible to detect a solar related responses in station-based or reanalysis-based DTR datasets related to a hypothesized CR–cloud link, as potential signals would be drowned in noise.

## 1. Introduction

It has been suggested that the diurnal range of surface-level air temperature (DTR) provides an indirect means of testing for the presence of a solar–climate link mediated by cloud cover [*Dragić et al*., 2011]. This is an intriguing possibility, as large scale measurements of clouds can be prone to errors: the only viable means of large-scale measurement is via satellite-based detections, which are known to have numerous associated issues, including long-term instrumentation drifts, calibration errors, and view-angle artifacts [*Campbell*, 2004; *Norris*, 2005; *Pallé*, 2005; *Evan et al*., 2007]. Consequently, the idea proposed by *Dragić et al*. [2011] that the DTR may provide an indirect means of observing significant short term variations in cloud connected to solar activity is worth investigating further. There may potentially be great advantage to such an approach, as surface level air temperatures are far simpler to measure than cloud properties and the temperature records extend over a considerably longer time period.

Historical records of DTR over the 20th century suggest that as much as 80% of its long-term variance over large regions of the globe can be explained by an inverse relationship to a combination of cloud and precipitation changes [*Dai et al.*, 1999]. It is known that clouds largely determine the geographic patterns of DTR, and that over land the presence of clouds and secondary effects from soil moisture and precipitation can result in DTR reductions of 25–50% relative to clear-sky days [*Dai et al.*, 1999].

Physically, an anti-correlation between DTR and clouds exists as optically thick low-level clouds effectively reflect incident shortwave (SW) solar radiation, reducing daytime maximum temperatures. This effect diminishes for clouds of increasing altitude. Clouds also absorb and re-emit outgoing terrestrial longwave (LW) radiation, resulting in warmer nighttime temperatures relative to clear-sky conditions. Low level clouds are generally more effective at reducing the SW radiation received by the surface as they have a high optical depth, and also re-emit LW radiation at higher temperatures (relative to higher altitude clouds), making them more effective at modulating the DTR than higher optically thinner clouds. The SW blocking effects have been identified as more important to modifying DTR than the re-emission of LW radiation [*Dai et al.*, 1999]. A clear example of the efficacy of relatively small cloud changes inducing appreciable DTR variations is shown by *Travis et al.* [2002], who note an approximately 1.8°C change in DTR over the United States of America due only to the absence of Jet condensation trails.

## 2. Datasets

In this work DTR is calculated from two datasets: 1) DTR is estimated over all land regions between 60°N–60°S from 6-hourly averaged NCEP/NCAR surface level air temperature reanalysis data [*Kalnay et al.*, 1996]. This data is available over a 2.5° × 2.5° resolution grid, over the time period of 1950–2012. Due to the use of 6-hour averaged data, our estimation of DTR is more muted than it would be from the use of a more frequently sampled dataset, however, we have the benefit of regular spatial/temporal sampling across the globe. 2) DTR is also estimated from 210 meteorological stations (surface data, global summary of day dataset), which include the 189 stations used by *Dragić et al*. [2011]. The DTR is calculated from the daily maximum and minimum temperature difference. These stations are located within a region between 77.7°N–34.7°N, 179.4°W–170.4°E. The data are from the World Data Center for Meteorology, Asheville, USA, and are maintained by NOAA (www.ncdc.noaa.gov/CDO/cdo).

To give an estimate of the difference between the two DTR datasets we provide a comparison of the seasonal DTR climatology in Figure 1 with the reanalysis data restricted to an identical region covered by the station data: the seasonal variations show a good agreement ($r = 0.92$). However, the station-based data changes in advance of the reanalysis data, and show a peak around June, whereas the reanalysis-based data show a peak around September. The station-based data show a difference of 2.6°C between JJA and DJF months, while the reanalysis estimate of DTR shows a change of 2.1°C over the same period: based on these values we assume that the reanalysis-based DTR values are underestimated by at least 19%. This underestimation is likely caused by the relatively low temporal resolution of the data. The two deseasonalized DTR daily datasets show a correlation of $r = 0.44$ over the 1951–2010 period. The reanalysis-based DTR data should be considered as indicative rather than definitive. It is more likely that a solar-related DTR signal may be detected in the station data rather than the reanalysis data, as the reanalysis is based on observational data combined with climate model simulations. As climate models do not include hypothesized link between the cosmic ray (CR) flux and climate, potential CR–cloud signals (including an influence on DTR) may be dampened or removed entirely, although indirect effects may still remain.

To evaluate the relationship between DTR and solar activity we have used daily average total solar irradiance (TSI) flux data from the Physikalisch-Meteorologisches Observatorium Davos (PMOD) World Radiation Centre composite [*Fröhlich and Lean*, 1998; *Fröhlich*, 2000] from 1978–2012, and combined CR flux data from the Moscow (55.47°N, 37.32°E, 200m, 2.43GeV) and Climax Colorado (39.37°N, 106.18°W, 3400m, 2.99GeV) neutron monitors from 1951–2012.

Throughout this work we have also compared the observed DTR variations to a calculated upper limit CR flux–DTR response that may be expected. This upper limit is estimated from case 2 of *Dai et al*. [1999], who show that a roughly 70% increase in cloud results in a DTR reduction of approximately 3.6°C during summer months, and from *Svensmark et al*. [2009], who estimate that for an 18% reduction in the CR flux (measured from Climax) low cloud cover was reduced by 1.7%. From these values we estimate the following upper limit relationship: a 1% change in the CR flux may at most induce an anti-correlated DTR variation of 0.005°C. We note that the results of both studies may not be globally applicable, and also note that the findings of the later study have been criticized by *Laken et al*. [2009] and *Čalogović et al*. [2010]. Consequently, while the upper limit is almost certainly overestimated, it provides an indicative point of comparison enabling us to gauge observations against theoretical maximum expectations under the most favorable CR–cloud link scenario.

It must also be noted that our analysis has focused on DTR over land as this is where the majority of observations exist, and also where the DTR is largest. However, it is predicted that oceanic environments with low ambient aerosol concentrations may be the most sensitive to a CR–cloud link, although no evidence of such a relationship has been identified in such locations at either short or long timescales using modern satellite cloud data [*Kristjánsson et al*., 2008; *Laken et al*., 2012].

**3. DTR and solar cycles**

The deseasonalized DTR (°C), the CR flux (%), and the TSI flux (W/m$^2$) are presented in Figure 2 over the 1951–2012 period. These data span more than five complete 11-year solar cycles. To investigate if any significant periodic/quasi-periodic variations exist in the DTR connected to solar periodicities we employed a period analysis, discussed in the following section.

*3.1. Lomb-Scargle period analysis*

The CR flux, reanalysis-based DTR and station-based DTR datasets have 14146, 22280, and 11987 unevenly sampled data points respectively. These data were binned in to 10-day periods and for each bin the mean and root mean square (RMS) error values were calculated. For the CR flux data the 11-year solar cycle was removed by subtracting a heavily smoothed version of the data (smoothed with a width of 501 days). To search for periodic/quasi-periodic modulations we use the method of Lomb-Scargle to compute the periodograms [*Scargle*, 1982; *Press et al.*, 1992]. We used the constraints imposed by the Nyquist frequency and the typical duration of the dataset to limit the range of different frequencies searched, and the number of independent frequencies was determined using the method of *Horne and Baliunas* [1986]. The resulting periodograms are presented in Figure 3.

For the reanalysis-based DTR and station-based DTR datasets, both periodograms show the strongest peak at 0.002738cycles/day (365.25-day) along with harmonic frequencies (Fig. 3a–b). This shows that despite de-seasonalization of the data prior to the period analysis, residuals of the seasonal frequency remain, and these features are the only statistically significant periods detected in the dataset. No other obvious peaks are present, clearly demonstrating that there is no association between the 11-year solar cycle and DTR variations.

*3.2 DTR and the 1.68-year cosmic ray period*

In the CR flux periodogram (Fig. 3c) the highest peak in the spectrum after the removal of the 11-year solar cycle appears at a frequency of 0.001579cycles/day (corresponding to a period of 633.3-day). The coherence, defined as the center frequency divided by the FWHM width of the peak, is 320 and is consistent with a periodic modulation. The amplitude of this modulation is found to be 0.6%.

The standard procedures for detecting periodic features in a periodogram are based on estimating the noise spectrum and using this to define the point above which we are unlikely to observe a random fluctuation in the periodogram. In other words, one determines the significance of possible peaks above the noise level. The standard false alarm probability estimate from the Lomb-Scargle algorithm gives the significance of the highest peak in the power spectrum assuming that all the data points are independent. However, in the presence of correlated data (i.e. red-noise), we will have to take a different approach in order to properly estimate the significance of the peaks evident in the periodogram. This was done numerically by means of Monte Carlo (MC) simulations. We generated data with exactly the same sampling as the real data with a model red-noise data generated using the method of *Timmer and König* [1995] with a broken power-law as determined from the periodogram of the observed data, and then added Gaussian noise using the uncertainties in data (we used the RMS value from the 10-day bin as an estimate, which is most likely too large). We then calculate the Lomb-Scargle periodogram and record the position and frequency of the highest peak. We computed 5,000 simulated datasets and then calculated the 68% and 99.9% confidence levels at each frequency taking into account a realistic number of independent trials [*Vaughan*, 2005]. In Figure 3b we show these confidence levels for the CR flux data, and it is clear that the peak in the periodogram corresponding to a 633.3-day period (0.001579cycles/day) is statistically significant at the 99.9% level.

This cycle has been noted by several authors as a 1.68-year CR flux periodicity [*Valdés-Galicia et al.*, 1996; *Valdés-Galica and Mendoza*, 1998; *Rouilliard and Lockwood*, 2004] and it is known to be connected to variations in the open solar magnetic flux. It has been previously used to examine solar climate links [*Harrison et al.*, 2011]. We note that although we detect a peak at 633.3-days (1.735-year), the 1.68-year period lies well within our uncertainty range, and we continue under the assumption that this is the same cycle.

The presence of a unique cycle in the CR flux lacking in the TSI flux provides an opportunity to disambiguate between these two inherently linked solar parameters over long timescales. However, we note that it is plausible that a 1.68-year DTR cycle may be difficult to detect, as the period is close to two-years in length, and may therefore may be masked by seasonal variations to an extent. This problem is further exacerbated as the cycle is low amplitude. The 1.68-year period has an amplitude of only 0.6% of the total CR flux: for comparison, the 11-year solar cycle possesses an amplitude of approximately 10%. We find no indication of the 633.3-day (1.68-year) solar cycle present in the DTR data. It is expected, that any potential DTR response resulting from the 1.68-year period may be too small to be detected, as our estimation of an upper limit DTR response suggests a change of 0.003°C at the most, which would not be significantly distinguishable from background noise if it is detected at all. However, assuming a physical CR–DTR relationship, we may expect to observe a DTR modulation corresponding to the 11-year solar cycle frequency. The lack of this cycle suggests no relationship exists between DTR and solar activity.

**4. DTR variations and Forbush Decreases**

An epoch-superpositional (composite) based analysis of the high magnitude Forbush decreases events (defined as an abrupt reduction of at least 3% in background CR intensity) has often been employed by investigators seeking to test the hypothesized solar–cloud link [e.g. *Pudovkin and Veretenenko*, 1995; *Todd and Kniveton*, 2001; *Kristjánsson et al.*, 2008, *Svensmark et al.*, 2009; *Harrison and Ambaum*, 2010; *Laken and Čalogović*, 2011]. This method of analysis is of particular use as it prevents interference from climate oscillations and periodicities (such as El Niño), which have been noted to influence long-term studies [*Farrar*, 2000; *Laken et al.* 2012]. It also overcomes issues of instrumental errors and calibration issues associated with long-term measurements [*Campbell*, 2004; *Norris*, 2005; *Evan et al.*, 2007]. Such factors are particularly problematic for satellite-based cloud measurements, and have resulted in uncertainties regarding a hypothesized link between solar activity and cloud cover. Despite the benefits of composite based analysis, it has also been found that Forbush decrease (FD) investigations may fail to distinguish the influences of the TSI and CR flux [*Laken et al.*, 2011], or be limited by high levels of variability inherent in small sample sizes to the point where reliable and significant detection of a solar signal in climatological datasets becomes unlikely [*Laken and Čalogović*, 2011].

We present an analysis of DTR variations occurring in association with FD events using 267 FD events taken from *Laken et al.* [2011]: these events are primarily based on the National Geophysical Data Center (NGDC) detected by Mt. Washington neutron monitor (NM) data (the same events as used in *Dragić et al.* [2011]), extended with additional data (the events used by *Dragić et al.* [2011] only extend to 1995). The FD events have been adjusted from the original NGCD data, altering the key composite date to reflect the date of maximal CR flux reduction rather than the FD onset date. This is required as the onset date can deviate by a period of hours to several days from the maximal CR flux deviation, resulting in significant complications to the composite analysis, as noted by *Troshichev et al.* [2008]; see *Laken et al.* [2011] for a full description of the date adjustment.

As a further test, we also select the events with the highest magnitude CR flux reductions (of ≥7%), to examine if a stronger CR flux reduction alters the results. Since the majority of the FD events are of low magnitudes, this approach reduces the sample size (from *n* 267 to *n* 29 events). *Dragić et al.* [2011] find that significant variations in DTR are detected only during high magnitude events (of ≥7% reductions in the CR flux). The values of our FD reductions appear somewhat smaller than the 7% threshold of *Dragić et al.* [2011] (Fig. 4h), this is because although our data are largely based on the same list of FD events (which have been adjusted and extended), our NM data was from a different source with a slightly different cut off rigidity (Rc): the *Dragić et al.* [2011] values were from Mt. Washington NM (Rc 1.46GeV), whereas our data are from a combination of Moscow NM (Rc 2.46GeV), and Climax Colorado NM (Rc 3.03GeV). Additionally, the magnitude of the *Dragić et al.* [2011] FD events were calculated from the increase in CR flux preceding and FD event, to the peak reduction, whereas our values are calculated as an anomaly against a 21-day averaging period.

Throughout this analysis we use MC based techniques to assess statistical significance. This method of significance testing has great advantages over alternative testing methods (such as the Student's t-test) as it makes no assumptions of the dataset and is consequently very robust.

*4.1. Results of Forbush Decrease based analysis*

Figure 4 shows the results of the FD analysis over a ±40 day period surrounding the key date of CR decrease for the global (NCEP/NCAR reanalysis) DTR data, regional (station-based) DTR data, the CR flux and the TSI flux. All data are differenced against a 21-day moving average period to remove long-to-middle-term variability. This procedure has not affected the sensitivity of the experiment, as changes associated with FD events (including the hypothesized cloud responses) occur within an approximately one-week period (e.g. Fig. 4g–h show that the CR reductions are clearly isolated and strongly significant with this method).

For both samples (*n* 276 and *n* 29) a clear reduction in the CR flux and the TSI flux is seen around the key composite date, however there are no corresponding significant variations in DTR at either global or regional scales. Due to smaller samples sizes in the case of the higher magnitude FD events all datasets show larger levels of variability. However, the largest variations in DTR occur before either the TSI or CR flux reductions have begun and are clearly not connected to solar factors. Our roughly estimated upper limit of a CR-related DTR response in the composites for *n* 276 and *n* 29 is found to be 0.016°C, and 0.061°C respectively. In both instances, the upper limits estimated from a favorable CR–cloud link scenario are much smaller than the noise present in the reanalysis and station data. Consequently, the FD analysis can not exclude the possibility of a CR–cloud link, but rather, it is possible to calculate an an upper limit value to detect a CR–DTR effect (at the 95% confidence level). This is done by dividing the MC-calculated confidence intervals by the CR reduction of the composite samples: For the *n* 276 sample, the two-tailed 95% confidence interval is found to be ±0.07°C (Fig. 4c), and the peak CR flux reduction is 2.41%, giving an upper limit to a CR–DTR effect (at the 95% confidence) of 0.03°C. For the *n* 29 sample, the confidence interval is found to be 0.22°C (Fig. 4d), while the peak CR flux reduction is 6.30%, giving an upper limit to a CR–DTR effect (at the 95% confidence) of 0.035°C. Thus, from the DTR composites we may conclude that if the CR flux is affecting the DTR, this relationship is no larger than approximately 0.03°C per 1% change in CR flux (constrained by the noise in the DTR data), and this upper limit is well above the expected signal (of at most 0.005°C per 1% change in CR flux).

Figure 5a–b shows the DTR variations occurring at a local (2.5° × 2.5°) resolution on day +3 and +6 of the high magnitude FD (≥7%) event composite. Locally, DTR variations of as much as ±2.5°C are observed. The spatial extent and magnitude of the positive and negative DTR variations are approximately equal. Over the 3-day period presented the DTR variations show large differences, implying that the variations are predominately stochastic. To assess if any abnormal changes in the number of statistically significant pixels occurred around the key composite date (day 0) we have examined the field significance (total percentage of pixels identified as locally significant on each day) over a ±40 day period (Fig. 5c). The statistical significance of each pixel was established based on the 95th percentile (two-tailed) DTR variations extracted from 10,000 MC simulated composites for individual pixels of data. We observed no abnormal variations in field significance prior to or following the FD events, indicating no unusual changes in DTR variations are occurring at local scales.

To investigate if significant local scale DTR variations are occurring in the station-based data, we converted the data into a gridded dataset through Kriging interpolation [*Isaaks and Srivastava*, 1990]. The key date interpolated DTR anomalies for the ≥7% FD composite are presented in Figure 6 (where in this instance anomaly is defined as the difference between the date in question and a five-day averaging period beginning seven days earlier): although increases in DTR can be seen over areas of Northern Europe and central Russia, decreases in DTR of similar magnitudes and spatial extent are also observed over regions of Eastern Europe and Siberia, again indicating the stochastic nature of the results. Furthermore, none of the anomalies are found to be significant above the 95th percentile (two-tailed) level, and no significant anomalies were identified between day +1 and +6.

*4.2. Reproducing the results of Dragić et al. [2011]*

We note that the results presented thus far have disagreed with the findings of *Dragić et al.* [2011], who claimed to identify a statistically significant increase in DTR following the largest magnitude FD onset dates. These authors calculated changes in DTR as deviations from expected climatological values; the expected values were calculated from a cubic spline fit to the data from 189 European region meteorological stations (for which we obtained identical and additional datasets). The dates of FD onset were selected by *Dragić et al.* [2011] for CR flux reductions greater than 7%, and were separated by at least a ±10 day period from other FD events or ground level enhancements (see Table 1). Note that the only difference between these dates and the dates previously used for the ≥7% sample is that these dates are for the onset of FD conditions, not the maximal CR flux reduction, and consequently isolate the CR/TSI flux reduction to a smaller extent than shown in our composite analysis presented so far.

Based on these FD onset dates we have re-analyzed the results presented in Figure 3 of *Dragić et al.* [2011]. The original figure is reproduced in Figure 7a. Using data from 210 meteorological stations we have calculated the DTR deviations on each day by subtracting a 21-day moving average from the data. These results are shown in Figure 7b over an extended (±40 day) time period: we indeed identify a DTR increase of 0.39°C between day 0 and +3, in close agreement to the findings of *Dragić et al.* [2011]. However, we find that the appearance of an unusual DTR increase presented by *Dragić et al.* [2011] is illusory: although we also find the day +3 DTR peak to be weakly significant based on an analysis of distributions of 100,000 MC-generated

DTR composites, a view of the data over a ±40 day period reveals these changes to be unremarkable (Fig. 7b): the peak is indistinguishable from numerous other variations of similar magnitude seen to violate the 95th percentile confidence interval over the 81-day period (as expected over a time period of this length). The restriction to a short (±10 day) period gives the false impression of an abnormal change in DTR. Furthermore, the observed CR anomalies (calculated from combined data from Climax and Moscow neutron monitors) show a peak reduction of 4.22% on day 4 of the composite, from which we estimate an upper limit DTR response of 0.02°C. The anomaly identified on day 3 of the *Dragić et al.* [2011] composite is far in excess of the predicted upper limit DTR response. Consequently, it is more likely that this result is caused by noise than a CR-related response.

**5. Conclusions**

From examinations of the 11-year and 1.68-year solar cycles, no evidence is found for a link between DTR and solar activity over 60 years of continuous data. However, with regards to the 1.68-year cycle we note that the detection of this periodicity in the DTR data may not be possible due to the low amplitude change in the CR flux (0.6%) over this cycle, although the additional lack of a significant DTR–CR relationship with the 11-year solar cycle suggests a DTR–solar link is unlikely.

Composite samples of FD events, corresponding to significant reductions in the TSI flux, also show no detectable relationship to DTR variations. Estimates of the noise in the DTR composites obtained with Mote Carlo simulations, suggest (with a 95% confidence) that if such a relationship does exist it is smaller than 0.03°C per 1% CR flux change (based on the NM data from Moscow and Climax). On the other hand, based on the most favorable CR–cloud response scenarios we expect the DTR to be no larger than 0.005°C per 1% CR flux change. In addition, we find that the reported correspondence between FD events and changes in the DTR by *Dragić et al.* [2011] were due to an overestimation of the statistical confidence of their result. In conclusion, we find no evidence to support claims of a link between DTR and solar activity. If this relationship exists, it is well within the meteorological noise of the datasets.

Thus, we conclude that previous claims of a significant DTR–solar response were the result of random DTR variability, and furthermore, the detection of any solar related responses in station-based and reanalysis-based DTR datasets via a hypothesised CR–cloud link is very likely not possible, as potential signals would be drowned in the noise.

**Acknowledgements**
The authors kindly thank Alex Dragić (University of Belgrade) and Juan Betancort (Instituto de Astrofísica de Canarias) for helpful correspondence, and three anonymous reviewers for the constructive comments. We acknowledge the NCEP Reanalysis Project data, provided by the NOAA/OAR/ERSL PSD, Boulder, Colorado, USA (www.cdc.noaa.gov). Meteorological station data are obtained from the World Data Center for Meteorology, Asheville, USA, maintained by NOAA (www.ncdc.noaa.gov/CDO/cdo). Cosmic ray data were obtained from the Solar Terrestrial physics division of IZMIRAN from http://helios.izmiran.rssi.ru. The authors acknowledges the PMOD dataset (version d41_62_1102): PMOD/WRC, Davos, Switzerland), which also comprises unpublished data from the VIRGO experiment on the ESA/NASA mission SoHO. FD event list obtained from http://www.ngdc.noaa.gov/stp/solar/cosmic.html. The authors acknowledge the European COST Action ES1005, and acknowledge support from the Spanish MICIIN, grant #CGL2009-10641 and AYA2010_18080.

**References**

Čalogović, J., C. Albert, F. Arnold, J. Beer, L. Desorgher, and E. O. Flueckiger (2010), Sudden cosmic ray decreases: No change of global cloud cover, *Geophys. Res. Lett.*, *37*(3), L03802, doi:10.1029/2009GL041327.

Campbell, G. (2004), View angle dependence of cloudiness and the trend in ISCCP cloudiness, *Am. Meterol. Soc.*, Norfolk, Va., 20–23 Sept.

Dragić, A., I. Anicin, R. Banjanac, V. Udovicic, D. Jokovic, D. Maletic, and J. Puzovic (2011), Forbush decreases – clouds relation in the neutron monitor era, *Astrophys. Space. Sci. Trans.*, *7*, 315–318, doi:10.5194/astra-7-315-2011.

Dai, A., K. E. Trenberth, and T. R. Karl (1999), Effects of Clouds, Soil Moisture, Precipitation and Water Vapor on Diurnal Temperature Range, *J. Climate*, *12*(8), 2451–2473, doi:10.1175/1520-0442(1999)012<2451:EOCSMP>2.0.CO;2.

Evan, A., A. Heidinger, and D. Vimont (2007), Arguments against a physical long-term trend in global ISCCP cloud amounts, *Geophys. Res. Lett.*, *34*(4), L04701.1–5, doi:10.1029/2006GL028083.

Farrar, P.D. (2000), Are cosmic rays influencing oceanic cloud coverage – or is it only El Niño?, *Clim. Change*, *47*(1–2), 7–15, doi:10.1023/A:1005672825112.

Fröhlich, C (2000), Observations of irradiance variations, *Space. Sci. Rev.*, *94*(1–2), 15–24, doi:10.1023/A:1026765712084.

Fröhlich, C., and J. Lean (1998), The Sun's total irradiance: Cycles, trends and related climate change uncertainties since 1976, *Geophys. Res. Lett.*, *25*(23), 4377–4380, doi:10.1029/1998GL900157.

Harrison, R. G., and M. H. P. Ambaum (2010), Observing Forbush decreases in cloud at Shetland, *J. Atmos. Sol. Ter. Phys.*, *72*(18), 1408–1414, doi:10.1016/j.jastp.2010.09.025.

Harrison, R. G., M. H. P. Ambaum, and M. Lockwood (2011), Cloud base height and cosmic rays, *Proc. Roy. Soc. A.*, *467*(2134), 2777–2791, doi:10.1098/rspa.2011.0040.

Horne, J. H., and S. L. Baliunas (1986), A prescription for period analysis of unevenly sampled time series, *Astrophys. J.*, *302*(1), 757–763.

Isaaks, E. H., and R. M. Srivastava (1990), *An introduction to applied geostatistics*, Oxford University Press. New York.

Kalnay, E., M. Kanamitsu, R. Kistler, W. Collins, D. Deaven, L. Gandin, M. Iredell, S. Saha, G. White, J. Woollen, Y. Zhu, M. Chelliah, W. Higgins, I. Janowiak, K. Mo, C. Ropelewski, J. Wang, A. Leetmaa, R. Reynolds, R. Jenne, and D. Joseph (1996), The NCEP/NCAR 40-Year Reanalysis Project, *Bull. Am. Met. Soc.*, *77*, 437–471, doi:10.1175/1520-0477(1996)077.

Kristjánsson, J., C. Stjern, F. Stordal, A. Færaa, G. Myhre, and K. Jonasson (2008), Cosmic rays, cloud condensation nuclei and clouds – a reassessment using MODIS data, *Atmos. Chem. Phys.*, *8*(24), 7373–7387, doi:10.5194/acp-8-7373-2008.

Laken, B., and J. Čalogović (2011), Solar irradiance, cosmic rays and cloudiness over daily timescales, *Geophys. Res. Lett.*, 38, L24811, doi:10.1029/2011GL049764.

Laken, B., A. Wolfendale, and D. Kniveton (2009), Cosmic ray decreases and changes in the liquid water cloud fraction over the oceans, *Geophys. Res. Lett.*, *36*, L23803, doi:10.1029/2009GL040961.

Laken, B., D. Kniveton, and A. Wolfendale (2011), Forbush decreases, solar irradiance variations and anomalous cloud changes, *J. Geophys. Res.*, *116*, D09201, doi:10.1029/2010JD014900.

Laken, B., E. Pallé, and H. Miyahara (2012), A decade of the Moderate Resolution Imaging Spectroradiometer: is a solar – cloud link detectable?, *J. Clim.*, doi: 10.1175/JCLI-D-11-00306.1.

Norris, J.R. (2005), Multidecadal changes in near-global cloud cover and estimated cloud cover radiative forcing, *J. Geophys. Res.*, *110*, D08206, doi:10.1029/2004JD005600.

Pallé, E. (2005), Possible satellite perspective effects on the reported correlations between solar activity and clouds, *Geophys. Res. Lett.*, *32*(2), L03802, doi:10.1029/2004GL021167.

Press, W.H., S.A. Teukolsky, W.T. Vetterling, and B.P. Flannery (1992), *Numerical recipes in C: The art of scientific computing, Second Edition*, New York: Cambridge University Press.

Pudovkin, M.I., and S.N. Veretenenko (1995), Cloudiness decreases associated with Forbush-decreases of galactic cosmic rays, *J. Atmos. Sol. Terr. Phys.*, *57*(11), 1349–1355, doi:10.1016/0021-9169(94)00109-2.

Rouillard, A., and M. Lockwood (2004), Oscillations in the open solar magnetic flux with a period of 1.68 years: imprint on galactic cosmic rays and implications for heliospheric shielding, *Annales. Geophysicae*, *22*(12), 4381–4395, doi:10.5194/angeo-22-4381-2004.

Scargle, J.D (1982), Studies in astronomical time series analysis. II – statistical aspects of spectral analysis of unevenly spaced data, *Astrophys. J.*, *263*, 835–853.

Svensmark, H., T. Bondo, and J. Svensmark (2009), Cosmic ray decreases affect atmospheric aerosols and clouds, *Geophys. Res. Lett.*, 36(15), L1501, doi:10.1029/2009GL038429.

Timmer, J., and M. König (1995), On generating power law noise, *Astron. Astrophys.*, *300*, 707–710.

Todd, M., and D. Kniveton (2001), Changes in cloud cover associated with Forbush decreases of galactic cosmic rays, *J. Geoophys. Res.*, *106*(D23), 32031–32041, doi:10.1029/2001JD00045.

Travis, D. J., A. M. Carelton, and R. G. Lauritsen (2002), Climatology: contrails reduce daily temperature range, *Nature*, *418*, 601, doi:10.1038/418601a.

Troshichev, O., V. Y. Vovk, and L. Egrova (2008), IMF-associated cloudiness above near-pole station Vostok: impact on wind regime in winter Antarctica, *J. Atmos. Sol-Terr. Phys.*, *70*(10), 1289–1300, doi:http://dx.doi.org/10.1016/j.jastp.2008.04.003.

Valdés-Galicia, J. F., B. Mendoza (1998), On the role of large-scale solar photospheric motions in the cosmic-ray 1.68-year intensity variation, *Solar. Phys.*, *178*(1), 183–191, doi:10.1023/A:1004922926205.

Valdés-Galicia, J. F., R. Peres-Enriquez, and J. A. Otaola (1996), The cosmic ray 1.68-year variation: a clue to understand the nature of the solar cycle?, *Solar. Phys.*, *167*(1–2), 409–417, doi:10.1007/BF00146349.

Vaughan, S. (2005), A simple test for periodic signals in red noise, *Astron. Astrophys.*, *431*(1), 391–403, doi:10.1051/0004-6361:20041453.

**Figure captions**

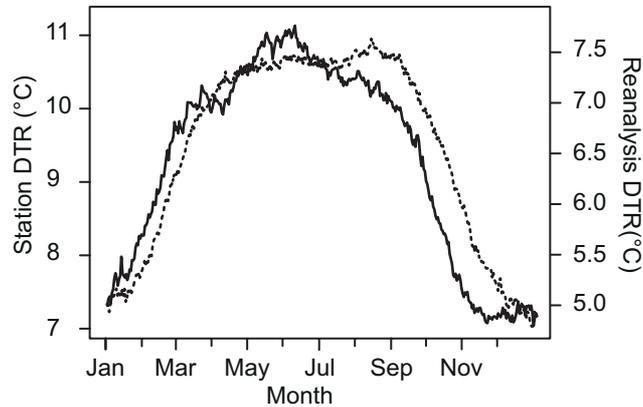

**Figure 1.**
Seasonal variations in DTR (°C) calculated from 210 meteorological stations data (solid line) over the region 77.7°N–34.7°N, 179.4°W–170.4°E and reanalysis data (dotted line) over an identical region (land-area pixels only), for the period of 1951–2010. Note the scale range of the left and right side *x*-axis for the purpose of comparison.

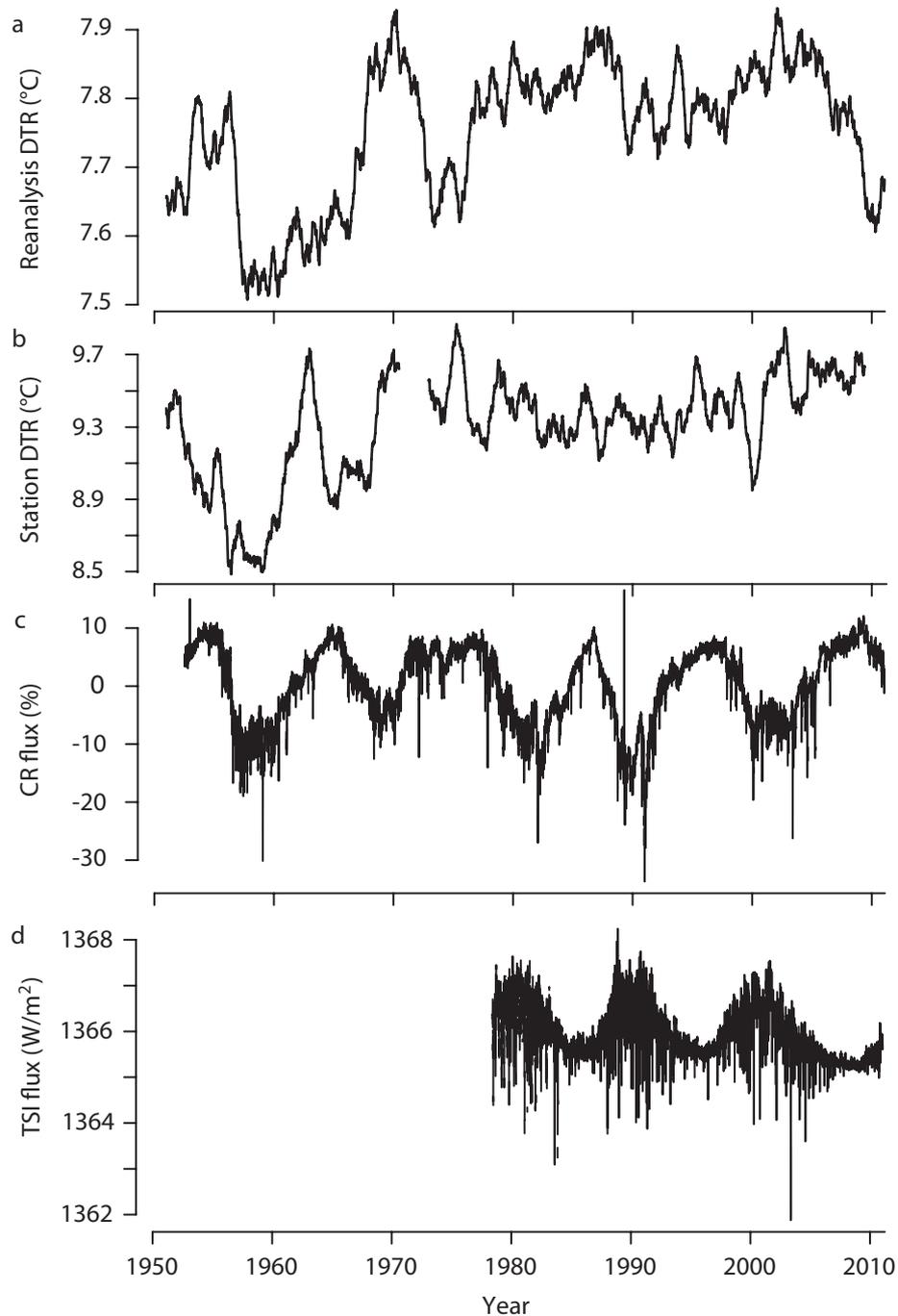

**Figure 2.**
a) De-seasonalized DTR calculated from reanalysis data over all land-area pixels from 60°N–60°S, b) de-seasonalised DTR calculated from 210 metrological stations, covering an area from 77.7°N–34.7°N, 179.4°W–170.4°E, c) the CR flux from the combined Climax Colorado and Moscow neutron monitor counts (%) normalized to 0% and d) the TSI flux from the PMOD reconstruction (W/m$^2$).

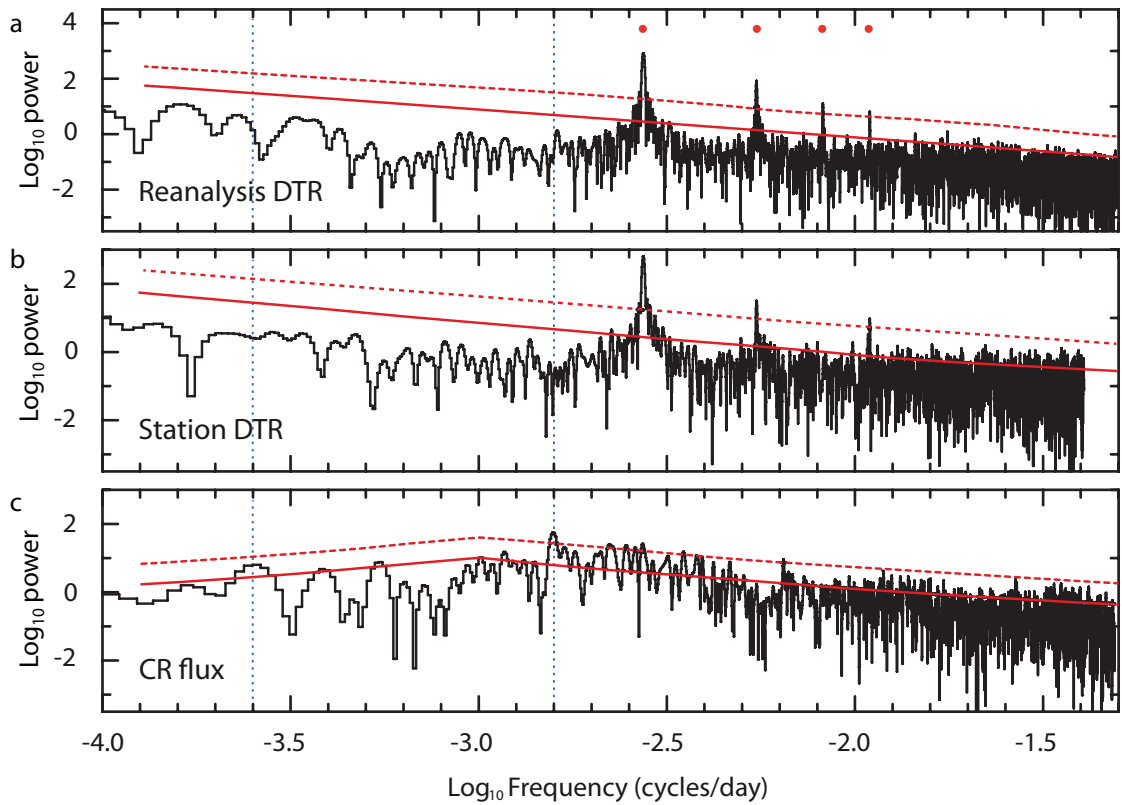

**Figure 3.**
Lomb-Scargle periodograms of de-seasonalised DTR from both a) reanalysis data, b) station data, and c) the CR flux with the 11-year solar cycle signal removed. Red markers indicate the 0.0027378cycles/day frequency (365.25 day) and its harmonics, clearly indicating the peaks in DTR are residual harmonics of the seasonal cycle, no other DTR peaks are significant. For the CR flux data the 68 percentile and 99.9 percentile red-noise confidence intervals at each frequency are indicated by the red solid and broken lines respectively from our MC analysis. As one can see frequency at 0.001579cycles/day (633.3 day) is significant at the 99.9 percentile level. Vertical blue dashed lines indicate the location of the 11-year solar cycle which has been removed as it would obscure other features of the periodogram, and the 633.3-day periodicity.

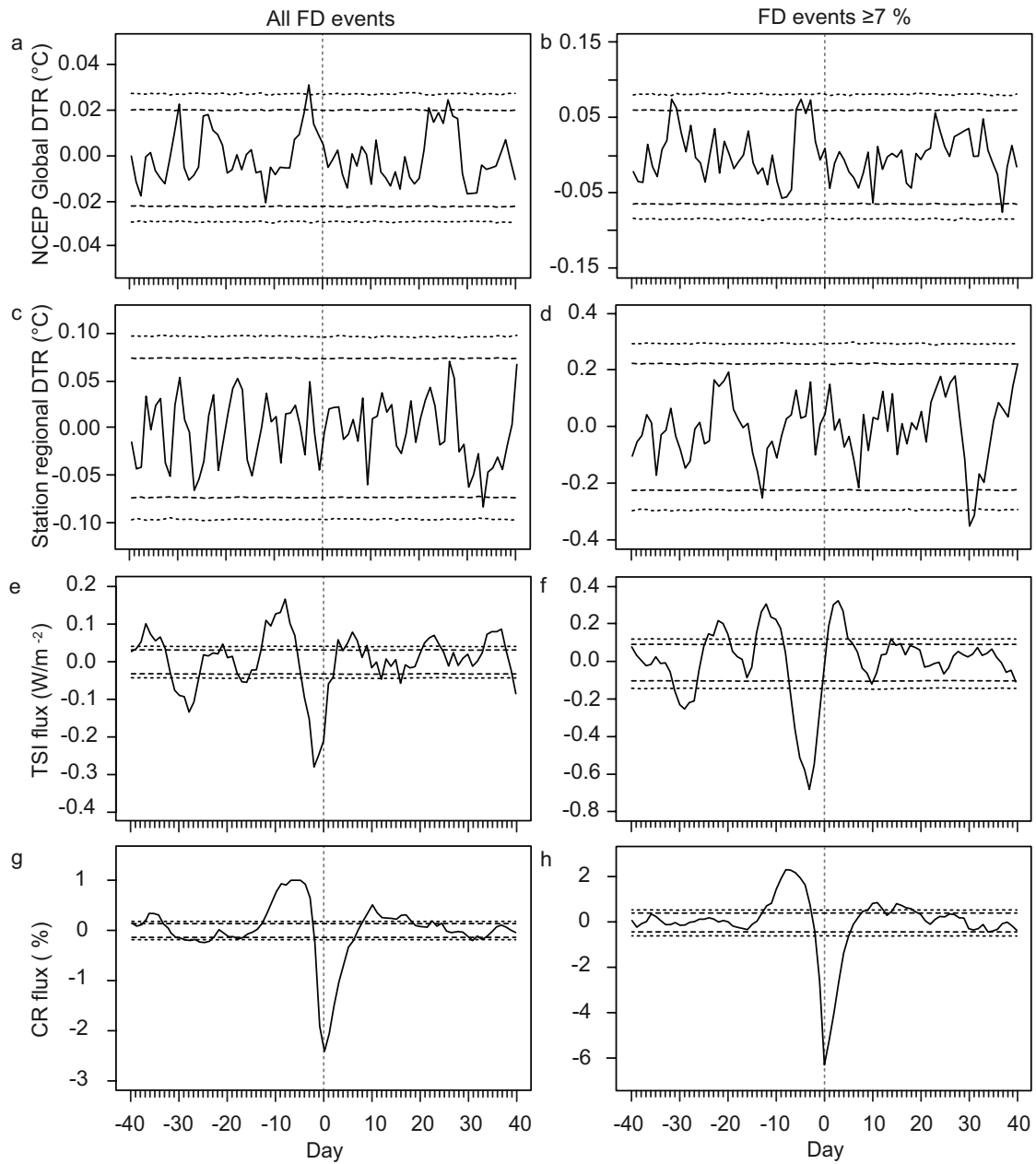

**Figure 4.**
Data presented over a ±40 day period. Right-hand-side plots show all FD events (*n* 267), while left-hand-side plots show only largest (≥7%) reductions in the CR flux (*n* 29). The dotted and dashed lines indicate the 99th and 95th percentile (two-tailed) confidence intervals respectively, calculated from 100,000 MC composites on each day of the composite. The key date (day 0) of the composite is indicated on each plot with a vertical grey dashed line and represents the day of maximal CR reduction. All values are residuals, calculated against a 21-day moving average. Panels a–b show global DTR from NCEP/NCAR reanalysis data (60°N–60°S, land-area pixels only); c–d show regional DTR from 210 meteorological stations located with a region of 77.7°N–34.7°N, 179.4°W–170.4°E; e–f show the TSI flux from the PMOD reconstruction; and g–h show the CR flux from the Climax/Moscow neutron monitors combined. The offset between TSI flux and CR flux minimum are due to the different transit times between irradiance and coronal mass ejections (which modulate the CR flux).

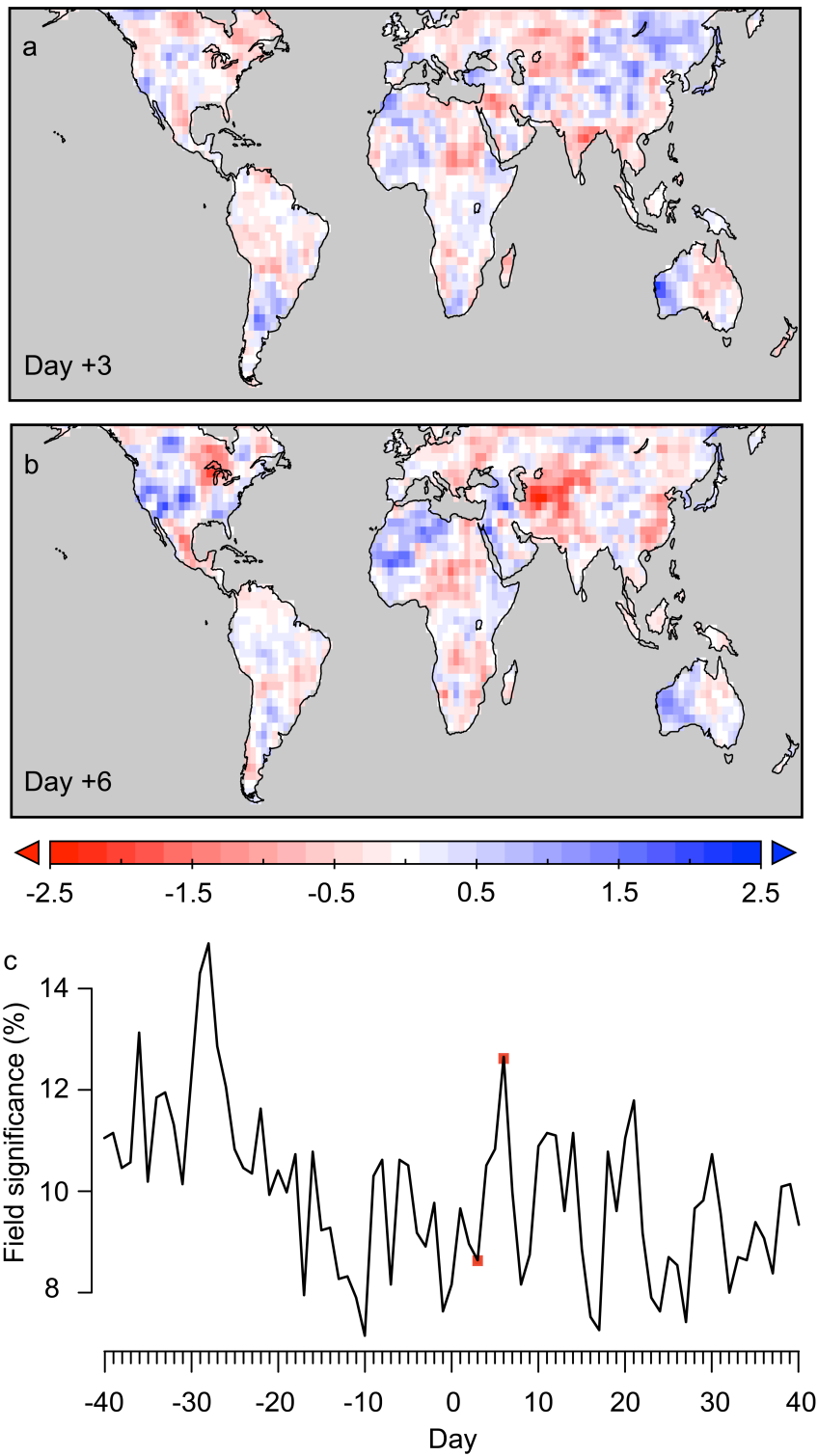

**Figure 5.**
DTR variations (over land-area grid cells only from 60°N–60°S) occurring on day 0 (key date) of the FD composites, for the largest (≥7 % CR flux reductions) FD events (*n* 29) on a) day +3, and b) day +6 of the composite. The percentage of locally significant pixels obtained each day over a ±40 day period surrounding the key date is presented in panel C. Red markers indicate the significance on day +3 and day +6. The statistical significance of each pixel is evaluated against a threshold value calculated from 10,000 MC generated DTR composites.

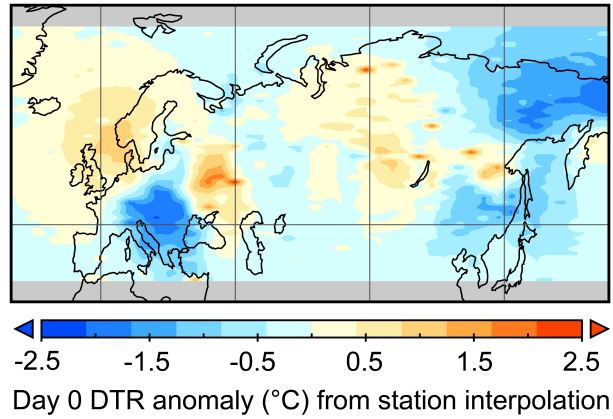

**Figure 6.**
Key date (day 0) DTR anomalies occurring during the ≤ -7% FD events, calculated from interpolated meteorological station data (using the Kriging method). The meteorological stations include the 189 stations used by *Dragić et al*. [2011]. No statistically significant anomalies were detected between days 0 to +6 (at the 95 percentile two-tailed level, calculated using MC methods). Anomalies are calculated against a five-day averaging period beginning on day -7.

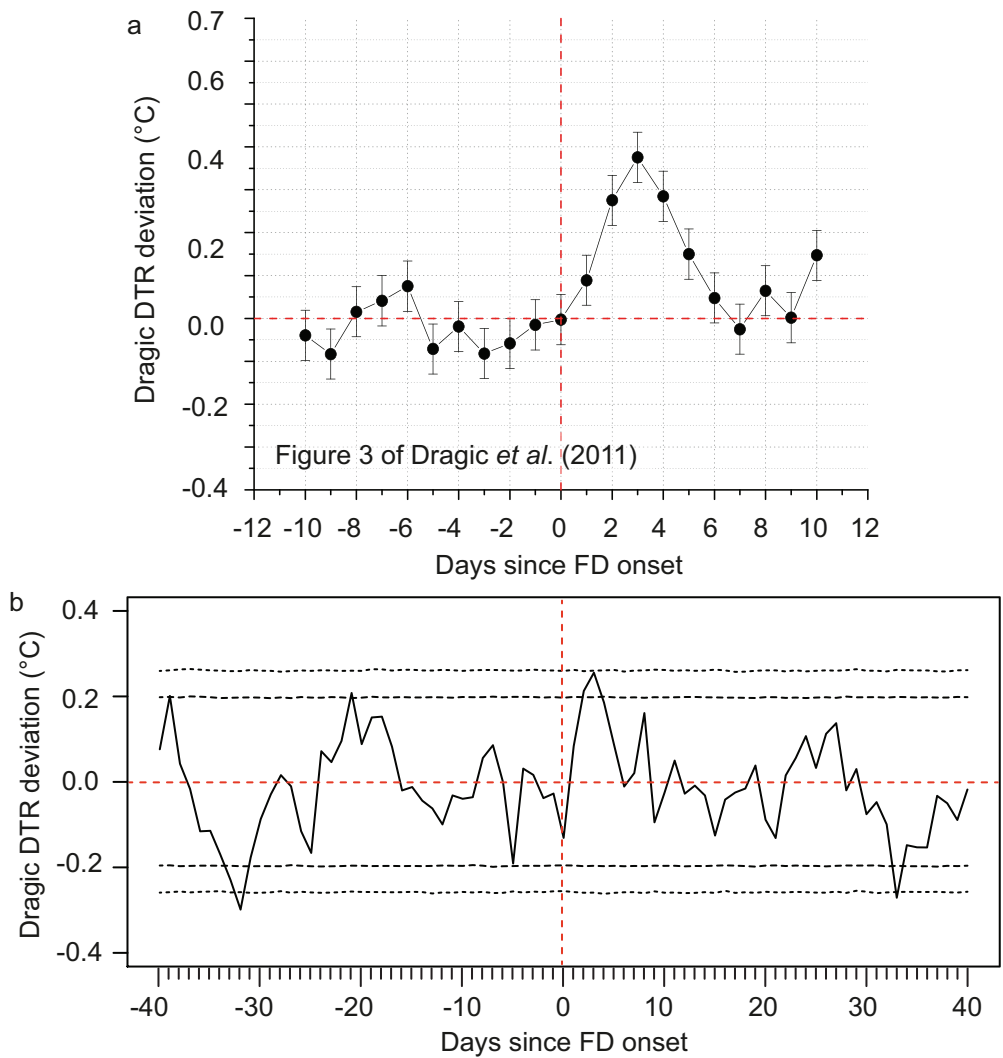

**Figure 7.**
a) Figure 3 taken from *Dragić et al.* [2011]. b) DTR deviations calculated from meteorological station data from 210 sites (between 77.7°N–34.7°N, 179.4°W–170.4°E), including the 189 sites used by *Dragić et al.* [2011]. Dotted and dashed lines indicate the 99th and 95th percentile (two-tailed) confidence intervals respectively, calculated for each day from MC-generated simulations of 100,000 DTR composites. Solid black line shows the mean DTR deviations over a ±40 day period surrounding the FD onset date. Values are residuals from a 21-day averaging period. Dates of the composite are given in Table 1 (FD onset for reductions of ≥7%, *n* 37).

| FD onset dates G.T. -7% (no SPE/FD ±10 days) | | |
|---|---|---|
| 08/10/1956 | 26/05/1967 | 23/07/1981 |
| 08/11/1956 | 28/02/1968 | 30/01/1982 |
| 21/01/1957 | 31/10/1972 | 06/06/1982 |
| 10/03/1957 | 30/03/1976 | 10/07/1982 |
| 29/08/1957 | 14/02/1978 | 18/09/1982 |
| 23/10/1957 | 06/03/1978 | 06/02/1986 |
| 23/11/1957 | 29/05/1978 | 12/03/1989 |
| 26/03/1958 | 13/07/1978 | 04/09/1989 |
| 13/02/1959 | 26/08/1978 | 07/04/1990 |
| 12/05/1959 | 07/07/1979 | 24/03/1991 |
| 30/03/1960 | 06/06/1980 | 28/05/1991 |
| 22/09/1963 | 30/11/1980 | 27/10/1991 |
| 30/08/1966 | 24/02/1981 | |

**Table 1.**

List of high magnitude (≥7% reduction) FD onset events recorded by the Mt. Washing observatory 1955–1995. Dates coincident within a ±10 day period of another FD (>7% reduction) event or a solar proton event (SPE) resulting in a ground level enhancement are excluded from the list. Both FD and SPE lists can be obtained from www.ndgc.noaa.goc/stp/solar/cosmic.html. One date is omitted due to lack of data (31/10/1972).